\documentstyle[prl,aps,floats,psfig]{revtex}

\begin{document}

\widetext
\draft
\twocolumn[\hsize\textwidth\columnwidth\hsize\csname @twocolumnfalse\endcsname
\title{Ionic Reactions in Two Dimensions with Disorder}

\author{
Jeong-Man Park$^{1,2}$ 
and 
Michael W. Deem$^1$
}
\address{
$^1$Chemical Engineering Department, University of
California, Los Angeles, CA  90095-1592\\
$^2$Department of Physics, The Catholic University, Seoul, Korea}

\maketitle

\begin{abstract}
We analyze the dynamics of the ion-dipole pairing reaction in
the two-dimensional
Coulomb gas in the presence of
disorder. Sufficiently singular
disorder forces the critical temperature of the 
Kosterlitz-Thouless-Berezinskii fixed point to be non-universal.  
This disorder leads to anomalous ion pairing kinetics with 
a continuously variable decay exponent.
Sufficiently strong disorder
eliminates the transition altogether.
For ions that are chemically reactive,
anomalous kinetics with 
a continuously variable decay exponent  
also occurs in the high-temperature
regime.
The Coulomb interaction inhibits
reactant segregation, and so the ionic $\mbox{A}^+ +\mbox{B}^- \to \emptyset$
 reaction behaves like
the nonionic $\mbox{A}+\mbox{A} \to \emptyset$ reaction.
\end{abstract}

]

\section{Introduction}
The two-dimensional Coulomb gas has been the subject of careful
attention since the elucidation of its low-temperature phase 
by Kosterlitz and Thouless  \cite{Kosterlitz} and Berezinskii 
\cite{Berezinskii}.
Above a critical value of the dimensionless temperature, the
system approximately obeys Debye-H{\"u}ckle statistics (as it
does for all temperatures in three dimensions).
Below the transition temperature, ions of opposite charge pair to form dipoles.
The temperature at which this metal-insulator transition occurs is
universal in the absence of disorder.
In the related superfluid system, this universality
corresponds to the universal jump discontinuity in superfluid
density (see \cite{nelsonII} for a review).

  The dynamics of the
Coulomb gas under an external field has been analyzed by phenomenological
extensions of the static Kosterlitz-Thouless argument
\cite{McCauley,Huberman,Myerson,Ambegaokar,AmbegaokarII,Halperin,AmbegaokarIII,Fisher,Dorsey,Minnhagen}.
Different scaling regimes were found, and these are now understood
to correspond to the cases of weak, slowly-varying 
or strong, rapidly-varying external fields \cite{Bormann}.
While the equilibrium properties of the two-dimensional Coulomb
gas have been established rigorously via field-theoretic analysis
of the Sine-Gordon Hamiltonian 
\cite{Ohta,OhtaII,Knops,Amit},
there has been to date no rigorous, field-theoretic model for
the ionic {\em dynamics}, near the low-temperature critical point or
otherwise.

The dynamics of the two-dimensional
reaction $\mbox{A}^+ +\mbox{B}^- 
\to\emptyset$, where $\mbox{A}^+$ and $\mbox{B}^-$ are
ions of opposite charge, has  been studied in the high-temperature
limit by 
scaling arguments and computer simulation.  
In the absence of Coulomb interaction, the $\mbox{A}$ and $\mbox{B}$ reactants
segregate. This segregation leads to the diffusion-limited decay
law $\langle c_{\rm A}(t) \rangle \sim [n_0 / (8 \pi^2 D t)]^{1/2}$
\cite{Deem2}.  Local charge neutrality enforced by the
Coulomb interaction inhibits this segregation
of the reactants, allowing for a faster decay law.
The charge density still decays
as a power law, $\langle c_{\rm A}(t) \rangle \sim a t^{-x}$. 
 The decay exponent, $x$, has
been observed in computer simulations to range from $0.79 \pm 0.04$ 
\cite{Huber} to $0.85 \pm 0.05$ \cite{Jang} to unity with logarithmic
corrections \cite{Yurke}.  
Scaling theories have been proposed that lead to values for the
decay exponent from $0.85$ \cite{Jang,Ginzburg} to unity
 \cite{Oshanin,Ispolatov}.  An approximate, self-consistent 
treatment of the classical reaction diffusion equations 
leads to the prediction that the decay exponent is unity \cite{GinzburgII},
 although
logarithmic corrections cannot be excluded due to the use of mean-field type
equations.

Studies of single ion diffusion in correlated disorder have shown
that for sufficiently long-ranged, disordered potential fields,
anomalous diffusion
occurs (see, for example, 
\cite{Fisher,Kravtsov1,Kravtsov2,Bouchaud1,Bouchaud2,Honkonen1,Honkonen2,Derkachov1,Derkachov2}).
Ionic disorder in two dimensions creates just such a potential
field,  $v({\bf x})$, that leads to anomalous diffusion.
As in previous work, we assume the potential to be
Gaussian, with zero mean and
correlation function $\chi_{vv}(r)$.
The appropriate form for the Fourier transform at long wavelengths is
$\hat \chi_{vv}({\bf k}) = \int d {\bf x} \exp(i {\bf k} \cdot {\bf x})
\chi_{vv}({\bf x}) = \gamma / k^2$.
This type of disorder leads to anomalous diffusion with a continuously
variable exponent
$\langle r^2(t) \rangle \sim b t^{1- \delta}$,
where $\delta = 1/[ 1 + 8 \pi/(\beta^2 \gamma)]$.

In this article, we use the rigorous field-theoretic formulation of 
reaction kinetics \cite{Peliti,Lee1,Lee2} to analyze both the ion-pairing
reaction near the metal-insulator transition and the 
$\mbox{A}^+ + \mbox{B}^- \to \emptyset$ annihilation
reaction at high temperatures.
The master equation description of this
reaction is described in Section II.  The field theory that we derive
from this description is presented in Section III.
Two dimensions is the upper critical dimension for this
system---the dimension below which mean field theory fails.
We derive the renormalization group flows for this system in
Section IV.
We give an asymptotically exact renormalization group 
analysis of the long-time dynamics in Section V.
For the low temperature phase, we 
find a decay exponent that depends continuously
on the strength of disorder.  Moreover, we find that the critical temperature,
which is universal in the absence of disorder, depends continuously on the
strength of disorder.  
In Section VI we analyze the high-temperature dynamics of the
$\mbox{A}^+ +\mbox{B}^-
 \to \emptyset $ chemical reaction.
We find a classical decay in the absence of disorder and
anomalous kinetics in the presence of disorder.  The Coulomb
interaction prevents segregation of the reactants under all
conditions, and so the dynamics of the ionic $\mbox{A}^+ +\mbox{B}^-
 \to \emptyset $ reaction
is similar to that of the neutral $\mbox{A}+\mbox{A}
 \to \emptyset $ reaction.
We conclude in section VII with a discussion of the
experimental implications of our results.

\section{Master Equation for Low-Temperature Ion Pairing}
To analyze the ion pairing that takes place below the transition
temperature, we consider the following reaction
\begin{equation}
\mbox{A}^+ + \mbox{B}^-
~{\mathrel{\mathop{\rightleftharpoons}\limits^{\lambda}_{\tau}}}~
 \mbox{C} \ ,
\label{1}
\end{equation}
where $\mbox{A}^+$ and $\mbox{B}^-$ are the ions of opposite charge, and
 $\mbox{C}$ is the
dipole.  We choose initially to have equal densities of ions
$\langle c_{\rm A}(0) \rangle = \langle c_{\rm B}(0) 
\rangle = n_0$ and no dipoles.
The ions are initially distributed at random, with Poissonian statistics.
The long-time decay is not sensitive to short-ranged correlations that might
be present in the initial conditions, such as those resulting from
a high-temperature quench.
  The ion-dipole interaction will prove to be irrelevant,
and so we can ignore the dipole orientation.
The presence of the dipoles will, however, be relevant, and so
it is necessary to include the reaction (\ref{1}).

By considering the reaction on a lattice, we can write
a master equation that governs changes in the
densities of $\mbox{A}^+$, $\mbox{B}^-$, and 
$\mbox{C}$.  The master equation relates how the probability, $P$, of a given configuration of
particles on the lattice changes with time:
\begin{eqnarray}
&&\frac{\partial P(\{ m_i \}, \{ n_i \}, \{ l_i \}, t) }{\partial t}  =
\nonumber \\  &&
\frac{D_{\rm A}}{(\Delta r)^2}
\sum_{i,j} [
T_{ji}^{\rm A}
 (m_j + 1) P(m_i-1, m_j+1, t)
\nonumber \\  &&
~~~~~~~~~~ -
T_{ij}^{\rm A}
m_i P ]
\nonumber \\  &&
+\frac{D_{\rm B}}{(\Delta r)^2}
\sum_{i,j} [
T_{ji}^{\rm B}
(n_j + 1) P( n_i-1, n_j+1, t)
\nonumber \\  &&
~~~~~~~~~~-
T_{ij}^{\rm B}
n_i P ]
\nonumber \\  &&
+\frac{D_{\rm C}}{(\Delta r)^2}
\sum_{i,j} [
(l_j + 1) P(l_i-1, l_j+1,  t)
- l_i P ]
\nonumber \\  &&
+ \frac{\lambda}{(\Delta r)^2} \sum_i [
(m_i+1)(n_i+1) P(m_i+1, n_i+1, l_i-1, t)
\nonumber \\
&&
~~~~~~~~~~
 - m_i n_i P
]
\nonumber \\
&&
+ \tau \sum_i [
(l_i+1) P(m_i-1, n_i-1, l_i+1, t)
 - l_i P ]
\label{1b}
\end{eqnarray}
Here $m_i$ is the number of A ions on site $i$,
$n_i$ is the number of B ions on site $i$, and $l_i$
is the number of C particles at site $i$.
The summation over $i$ is over all sites on the lattice,
and the summation over $j$ is over the nearest
neighbors of site $i$. The lattice
spacing is given by $\Delta r$.  The diffusive transition
matrix for hopping from site $i$ to a nearest neighbor site $j$
is given by $T_{ij}^{\rm A} = [1 + \beta (u_i^{\rm A} - u_j^{\rm A})/2]$ and
$T_{ij}^{\rm B} = [1 + \beta (u_i^{\rm B} - u_j^{\rm B})/2]$.  Here $u$ is the
sum of an external, quenched potential and the Coulomb
potential created by all of the {\em other} ions.   Specifically,
$u_i^{\rm A} = v_i + \sum_k [m_{i+k} - \delta_{k 0} - n_{i+k}] c_k$
and 
$u_i^{\rm B} = -v_i + \sum_k [n_{i+k} - \delta_{k 0} - m_{i+k}] c_k$.
Here $v_i$ is the external potential at site $i$, and
$c_k$ is the Coulomb interaction $c({\bf r}) = -J \ln (r) / (2 \pi)$.
For simplicity we will assume that the ions have the same diffusivity,
$D_{\rm A} = D_{\rm B} = D$.
The inverse temperature is given by $\beta = 1/(k_{\rm B} T)$.

\section{The Field Theory}
Using the coherent state representation, we map the master
equation onto a field theory \cite{Peliti,Lee1,Lee2}.
We incorporate a random potential into the field theory via the
replica trick \cite{Deem2,Kravtsov1}.  We must also incorporate
the ionic interaction into the field theory, taking care with
self interaction terms.

The field theory that we generate is quadratic in the fields
associated with the dipole density.  Integrating out these fields, we
are left with the action $S = S_0 + S_1 + S_2 + S_3 + S_4 + S_5$
\begin{eqnarray}
S_0 &=& \int d^d {\bf x} \int_0^{t_f} d t\,
  \bar a_\alpha({\bf x},t) \left[
\partial_t - D \nabla^2 + \delta(t)
 \right]
 a_\alpha({\bf x},t)
 \nonumber \\
&&+
\int d^d {\bf x} \int_0^{t_f} d t\,
\bar  b_\alpha({\bf x},t) \left[
\partial_t - D \nabla^2 + \delta(t)
 \right]
 b_\alpha({\bf x},t)
 \nonumber \\
S_1 &=& -n_0 \int d^d {\bf x}
\left[ \bar a_\alpha({\bf x},0) +
\bar b_\alpha({\bf x},0)  \right]
\nonumber \\
S_2 &=&
\lambda \int d^d {\bf x} \int_0^{t_f} d t
\left[
\bar a_\alpha({\bf x},t)
\bar b_\alpha({\bf x},t)
+\bar a_\alpha({\bf x},t)
+\bar b_\alpha({\bf x},t)
\right]
\nonumber \\
&&\times
 a_\alpha({\bf x},t)
 b_\alpha({\bf x},t)
\nonumber \\
S_3 &=& -\lambda \tau 
\int d^d {\bf x} d^d {\bf x}'
 \int_0^{t_f} d t d t'
\nonumber \\
&&
\times 
\left[ \bar a_\alpha({\bf x},t)
\bar b_\alpha({\bf x},t)
+\bar a_\alpha({\bf x},t)
+\bar b_\alpha({\bf x},t) \right]
\nonumber \\
&& \times
G\!\!\!\!/({\bf x}, {\bf x}' \vert t,t')
a_\alpha({\bf x}',t')
b_\alpha({\bf x}',t')
\nonumber \\
S_4 &=& \beta J
\int_0^{t_f} d t  \int_{{\bf k}_1 {\bf k}_2 {\bf k}_3 {\bf k}_4}
\nonumber \\
&& \times (2 \pi)^d \delta({\bf k}_1+{\bf k}_2+{\bf k}_3+{\bf k}_4)
\nonumber \\ &&\times
\left[
\hat{\bar a}_{\alpha}({\bf k}_1, t)
\hat{     a}_{\alpha}({\bf k}_2, t) -
\hat{\bar b}_{\alpha}({\bf k}_1, t)
\hat{     b}_{\alpha}({\bf k}_2, t)
\right]
\nonumber\\
&& \times
\left[
\hat{\bar a}_{\alpha}({\bf k}_3, t)
\hat{     a}_{\alpha}({\bf k}_4, t) -
\hat{\bar b}_{\alpha}({\bf k}_3, t)
\hat{     b}_{\alpha}({\bf k}_4, t)
\right]
\nonumber \\ &&\times
\frac{{\bf k}_1 \cdot ({\bf k}_1+{\bf k}_2)}
{\vert {\bf k}_1+{\bf k}_2 \vert ^2}
\nonumber \\
S_5 &=& \frac{\beta^2 D^2}{2}
\int_0^{t_f} d t_1 d t_2 \int_{{\bf k}_1 {\bf k}_2 {\bf k}_3 {\bf k}_4}
\nonumber \\
&& \times (2 \pi)^d \delta({\bf k}_1+{\bf k}_2+{\bf k}_3+{\bf k}_4)
\nonumber \\ &&\times
\left[
\hat{\bar a}_{\alpha_1}({\bf k}_1, t_1)
\hat{     a}_{\alpha_1}({\bf k}_2, t_1) -
\hat{\bar b}_{\alpha_2}({\bf k}_1, t_1)
\hat{     b}_{\alpha_2}({\bf k}_2, t_1)
\right]
\nonumber\\
&& \times
\left[
\hat{\bar a}_{\alpha_3}({\bf k}_3, t_2)
\hat{     a}_{\alpha_3}({\bf k}_4, t_2) -
\hat{\bar b}_{\alpha_4}({\bf k}_3, t_2)
\hat{     b}_{\alpha_4}({\bf k}_4, t_2)
\right]
\nonumber \\ &&\times
{\bf k}_1 \cdot ({\bf k}_1+{\bf k}_2)
{\bf k}_3 \cdot ({\bf k}_1+{\bf k}_2)
\hat\chi_{vv}(\vert {\bf k}_1+{\bf k}_2\vert)
\ .
\label{2}
\end{eqnarray}
Summation is implied over replica indices.  The notation
$\int_{\bf k}$ stands for $\int d^d {\bf k} / (2 \pi)^d$.
The upper time limit in the action is arbitrary as long as
it exceeds times for which we wish to make calculations.
The random, Poissonian initial condition is accounted for
by the term  $S_1$.  
  The forward reaction is captured by the term $S_2$.
The effective potential due to the dipoles is captured
by the term $S_3$.  The propagator of the dipoles is given by
\begin{equation}
\hat {G\!\!\!\!/}(k,t) = 
\left\{
\begin{array}{l}
\exp[-(D_{\rm C} k^2 + \tau)t],~ t>0 \\[.2in]
                            0, t \le 0
\end{array}  \right. \ ,
\label{2a}
\end{equation}
where $D_{\rm C}$ is the diffusion coefficient of the
dipoles.  At long times, the ion density will be much smaller
than the dipole density, and 
we can replace the instantaneous
dipole density with the average density.  This simplifies
the effective dipole term to
\begin{eqnarray}
S_3' &=& - n_{\rm c} \tau
\int d^d {\bf x}
 \int_0^{t_f} d t 
\nonumber \\
&&\times
\left[ \bar a_\alpha({\bf x},t)
\bar b_\alpha({\bf x},t)
+\bar a_\alpha({\bf x},t)
+\bar b_\alpha({\bf x},t) \right] \ ,
\label{2b}
\end{eqnarray}
with $n_{\rm c} = n_0 - \langle c_{\rm A}(t) \rangle \sim n_0$.
This modified  action is identical to that for the reaction
\begin{equation}
\mbox{A}^+ + \mbox{B}^-
~{\mathrel{\mathop{\rightleftharpoons}\limits^{\lambda}_{n_{\rm c} \tau}}}~
\emptyset \ .
\label{2c}
\end{equation}
This reaction can be recognized as the one
addressed by the usual Sine-Gordon model of the Coulomb
gas, with the equilibrium ionic density, $y$,  given at low densities by
$y^2 = n_{\rm c} \tau / \lambda$.
The flow equations for these two forms, $S_3$ 
and $S_3'$,  are, of course, equivalent.
The Coulomb interaction between the ions is captured by
the term $S_4$.  Note that the Coulomb coupling should be an
effective one, including a finite renormalization due to dipole
screening.
The effective potential due to the randomness is captured in
the term $S_5$.

The concentrations, averaged over initial conditions, are given by
\begin{eqnarray}
\langle c_{\rm A}({\bf x},t) \rangle &=&
 \lim_{N \to 0} \langle a({\bf x},t) \rangle \
 \nonumber \\
\langle c_{\rm B}({\bf x},t) \rangle &=&
 \lim_{N \to 0} \langle b({\bf x},t) \rangle \
\ ,
\label{1a}
\end{eqnarray}
where the average on the right hand side is taken with respect to  $\exp(-S)$.

\section{Renormalization Group Flows}
We use renormalization group theory to deduce the long-time scaling
of the ionic concentration.  The diagrams that we need to consider
are illustrated in Figure 1.
\begin{figure}[t]
\centering
\leavevmode
\psfig{file=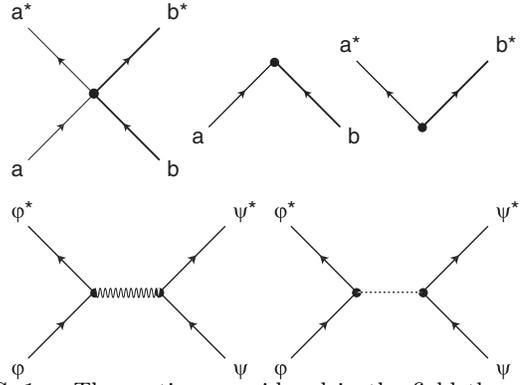,height=2in}
\caption[]
{\label{fig1}
The vertices considered in the field theory.  We have
set $a^* = \bar a + 1$ and $b^* = \bar b + 1$.  All combinations
of $\varphi = a,b$ and $\psi = a,b$ are considered.
}
\end{figure}
The one-loop
flow equations that result are
\begin{eqnarray}
\frac{d \ln n_{\rm c}}{d l} &=& 2
\nonumber \\
\frac{d \ln \lambda}{d l} &=& -
   \frac{\lambda}{4 \pi D} + \frac{\beta J}{2 \pi}
    - \frac{\beta^2 \gamma}{4 \pi}
\nonumber \\
\frac{d \ln \left( n_{\rm c} \tau / \lambda\right)}{d l} &=& 4
   - \frac{\beta J}{2 \pi}
    + \frac{\beta^2 \gamma}{2 \pi}
\nonumber \\
\frac{d \ln (\beta J)}{d l} &=& 
   - \frac{\lambda^2 \beta J}{2 \pi D^2}
     \frac{n_{\rm c} \tau }{\lambda \Lambda^4}
\nonumber \\
\frac{d \ln (\beta^2 \gamma)}{d l} &=& 
   - \frac{\lambda^2 \beta J}{\pi D^2}
     \frac{n_{\rm c} \tau }{\lambda \Lambda^4}
\ ,
\label{4}
\end{eqnarray}
where $\Lambda$ is the cutoff in Fourier space.
The dynamical exponent is given by
\begin{equation}
z = 2 + \frac{\beta^2 \gamma}{4 \pi}  + \frac{\beta^2 \gamma}{4 \pi} 
\frac{\lambda^2}{D^2} \frac{n_{\rm c} \tau }{\lambda \Lambda^4} \ .
\label{4a}
\end{equation}
These flow equations are valid to first order in $\tau$.  At this
order, they are valid to all orders in $\beta J$.
Also at this order,
the flow equation for $\beta^2 \gamma$ is likely valid
to all orders in $\beta^2 \gamma$ \cite{Deem2,Deem1}.  The flow
equation for $\lambda$ may contain contributions from
higher orders in $\beta^2 \gamma$.

\section{Matching and Results}

To compute the long-time value of the ionic concentration, we
integrate the flow equations up to a matching time, $t_0$.  We match
the results of the flow equations to a mean field theory that is
valid for short times.  At these short times, we need not worry about
renormalization of the reaction rates or Coulomb coupling.  Furthermore,
the reaction dynamics occurs in a local region, where the random
potential is roughly constant, and so we may assume
normal diffusive behavior.  In other words, we can use the standard, 
classical reaction diffusion equations.  A self-consistent treatment
of these equations has recently been presented \cite{GinzburgII}.
This theory suggests that the Coulomb interaction prevents
segregation of the reactants.  Moreover, the reaction is not limited
by local transport as long as $\lambda \le 2 \beta J D$.  We see that this
condition is satisfied by the fixed point forward rate, and so
the concentration is given  by
\begin{equation}
\langle c_{\rm A}(t(l),l) \rangle = 1/\left[1/n_0(l) + \lambda^* t(l) \right]
\ .
\label{7}
\end{equation}
We find the physical concentration from the relation 
\begin{equation}
\langle c_{\rm A}(t)\rangle = e^{-2 l} \langle c_{\rm A}(t(l), l) \rangle
\ .
\label{8}
\end{equation}
The result is
\begin{equation}
\langle c_{\rm A}(t)\rangle
 \sim \frac{1}{\lambda^* t} \left( \frac{t}{t_0} \right)^\delta \ ,
\label{9}
\end{equation}
with the fixed point reaction rate given from Eq.\ (\ref{4}) as
$\lambda^*/D = 2 \beta J^* - \beta^2 \gamma^* = 16 \pi + \beta^2 \gamma^*$.
Interestingly, we see that $\lambda^*$ is finite
at the 
Kosterlitz-Thouless fixed point, where $(n_{\rm c} \tau)^*$ vanishes.
Dipole dissociation, then, is key to the physics of the low-temperature
fixed point.

We see that the ions pair according to the 
classical law in the absence of disorder.  In the presence of
disorder, we find anomalous kinetics.  The kinetics is anomalous
because at long-times and low concentrations
the reaction becomes diffusion limited, and
at long times the diffusion is anomalous in the type of disorder
that we are considering.  Note that this result for the
ion pairing in disorder below the transition temperature is identical
to that for the $\mbox{A}+\mbox{A}
 \to \emptyset$ reaction with disorder \cite{Deem1},
except for a different value of $\lambda^*$.

If we interpret these flow equations as relations between related
{\em equilibrium} models, we recognize the standard
Kosterlitz-Thouless result when disorder is absent,
with $y^2 = \langle c_{\rm A} \rangle^2 = n_{\rm c} \tau / \lambda$.
 Figure 2 shows the
flows for the case of weak disorder.
\begin{figure}[t]
\centering
\leavevmode
\psfig{file=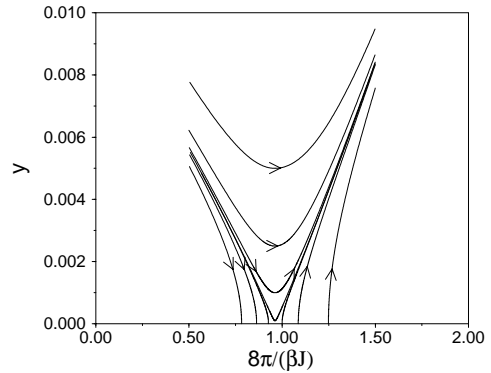,height=2in,angle=-90}
\caption[]
{\label{fig2}
The flow of the equilibrium ion density near the low-temperature critical
point.  Below 
the critical temperature
 the ion density is driven to zero as dipoles are
formed.  Above 
the critical temperature
the dipoles unpair, and the ion density becomes
large. 
For each curve, $\beta^2 \gamma = 1$ and $\lambda/D =
16 \pi + 1$ at the point closest to the
critical point.
}
\end{figure}
  Interestingly,
we can deduce the one-loop
critical temperature in the presence of
disorder with an extension of the
elementary Kosterlitz-Thouless free energy argument \cite{Kosterlitz}.
This argument predicts that the ion pairs will unbind when the free
energy to create two unbound ions, $F_\pm = (E_+ + E_-) - T (S_+ + S_-)$,  is
positive.  The Coulomb energy of the ion pair is, of course,
 $U_{\rm C}(r) = J \ln (r)
/ (2 \pi)$.  The effective interaction between an ion pair due to the
random potential is given by 
$U_{\rm V eff}(r) = -\beta^{-1}
\ln \langle \exp\{-\beta [v(0) - v(r)] \} \rangle  =
-\beta \gamma \ln (r) / (2 \pi)$.  
Quenched and annealed statistics are identical here for an ion pair
seperated by a finte distance, $r$, in 
a sufficiently large disordered medium, since the correlations in the
potential for the ion pair are short ranged \cite{Cates,Wu}.
The entropy  of the ion pair is, of course,
$4 k_{\rm B} \ln (r)$.  The ions, therefore, proliferate when 
$\beta J - \beta^2 \gamma < 8 \pi$.  This condition is exactly 
the one contained in the flows of Eq.\ (\ref{4})
near the low-temperature fixed point.   
This energy-entropy argument is not strictly rigorous, since the
metal-insulator transition occurs for $r \approx L$, where $L$ is the
system size.  In this regime, the correlations in the
potential for the ion pair are not short-ranged, and quenched and
annealed statistics are not strictly equal.  What we have shown
is that to one loop order these distinct statistics lead to the same
behavior.  Unless something unexpected occurs in the regime
$r \approx L$,  our location of the critical point may be exact
to all orders.

The transition temperature, which is universal in the absence of
disorder, becomes continuously variable in the presence of disorder.
This is a unique feature of the ionic disorder that we are considering.
The system undergoes a transition from
insulator to metal either by decreasing $\beta J$ or by
increasing the density of defects,
$\rho = \sqrt \gamma/J$.  Figure 3 shows the phase
diagram of the system at infinitesimally small
total (free plus bound)
ion density.
\begin{figure}[t]
\centering
\leavevmode
\psfig{file=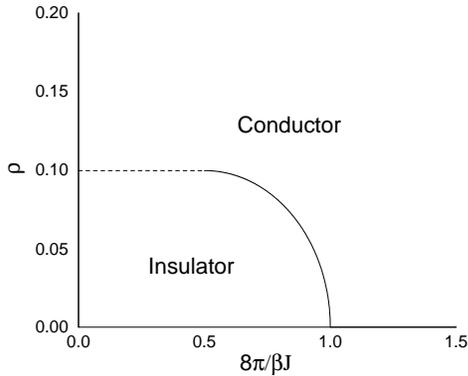,height=2in,angle=-90}
\caption[]
{\label{fig3}
The Kosterlitz-Thouless-Berezinskii 
fixed line in the presence of disorder (solid) for infinitesimally
small total ion density.  Here $\rho =
\sqrt \gamma/J$ is roughly the density of defects.
 The system is an insulator below the curve.
}
\end{figure}
 Note that sufficiently strong 
disorder eliminates the insulating phase completely.  A similar type of
equilibrium phase diagram has been predicted
for two-dimensional crystals with random
substitutional disorder \cite{Nelson2,Fertig}.  This 
substitutional disorder is equivalent, in our language, to
random, quenched dipoles.  So we see that quenched ions obeying
bulk charge neutrality behave in the long-wavelength
limit in the same way as random, quenched dipoles.

A reentrant metallic phase may occur at low temperatures.
This insulating to conducting transition
may occur because the forces arising from the disorder, which tend to
separate the ion pairs, are  a factor
$1/T$ greater than the bare Coulomb forces.  Figure 4 shows the
reentrant phase diagram predicted by the flow equations for
$\rho = \sqrt \gamma/J = 0.05$ for a 
range of initial values  of 
$y = \langle c_{\rm A} \rangle = [n_{\rm c} \tau / \lambda]^{1/2}$
 and $\beta J$.  
\begin{figure}[t]
\centering
\leavevmode
\psfig{file=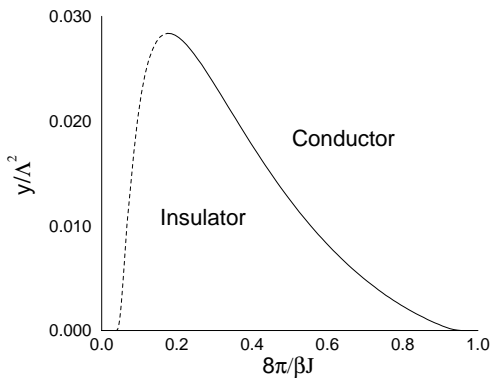,height=2in,angle=-90}
\caption[]
{\label{fig4}
The Kosterlitz-Thouless-Berezinskii 
fixed line in the presence of a fixed amount of
disorder, $\rho = \sqrt \gamma/J = 0.05$.  Note the reentrant phase
at low temperatures for a finite density of free ions, $y$.
The curve is strictly valid only in the high-temperature regime (solid).
}
\end{figure}
The temperature at which the reentrant phase occurs is roughly proportional
to $\sqrt \gamma$.  Since our flow equations are an expansion in
$\beta^2 \gamma$, they are not strictly valid in the reentrant regime.
Thus, the existence of
the reentrant phase, while physically plausible, cannot be rigorously
established with our flow equations.
A similar reentrant phase diagram has been predicted for the
equilibrium XY model with random Dzyaloshinskii-Moriya interactions,
which leads again, in our language, to a two-dimensional Coulomb gas with
random, quenched dipoles 
\cite{Shraiman}.

The ratio $\rho = \sqrt \gamma / J$ remains constant under renormalization.
This means that we can define
$\gamma = \gamma_0 / \epsilon^2$ and
$J = J_0 / \epsilon$, and one flow equation for $\epsilon$ will
result.  This factor
$\epsilon$ is none other than
the dielectric constant!  The disorder term contains two powers of the
dielectric constant because it is a correlation function of the
disorder potential.  The flow equation for $\epsilon$ is not
universal \cite{Ohta,OhtaII,Knops,Amit} and should probably include
additional (finite) terms.

\section{High-Temperature Dynamics}

We now turn to consider two-dimensional ionic reactions at high
temperatures.  That is, we consider the reaction
\begin{equation}
\mbox{A}^+ + \mbox{B}^-
~{\mathrel{\mathop{\to}\limits^{k}_{}}}~
 \mbox{P} \ ,
\label{11}
\end{equation}
where P is the neutral product of the reaction.
In the high-temperature regime, the ions pair to an insignificant
extent.  This follows from physical considerations.  This conclusion
also follows from the flow equations that drive the ion density to
large values.  Since the dipole density is insignificant, we may
ignore the ion-dipole reaction.  By comparing Eq.\ (\ref{11}) with
Eq.\ (\ref{2c}), we see that the appropriate action for this
reaction is Eq.\ (\ref{2}) with the replacement $\lambda \to k$,
$\tau \to 0$, and
$n_{\rm c} \to n_0$.
  The flow equations for this case are
\begin{eqnarray}
\frac{d \ln n_0}{d l} &=& 2
\nonumber \\
\frac{d \ln k}{d l} &=& -
   \frac{k}{4 \pi D} + \frac{\beta J}{2 \pi}
    - \frac{\beta^2 \gamma}{4 \pi}
\nonumber \\
\frac{d \ln (\beta J)}{d l} &=&  0
\nonumber \\
\frac{d \ln (\beta^2 \gamma)}{d l} &=&  0
\ .
\label{12}
\end{eqnarray}
In this case, $\beta J$ is a constant
to all orders.
As before \cite{Deem2,Deem1}, it seems likely that $\beta^2 \gamma$ is
a constant to all orders.   The flow equation for $k$ is accurate to
first order only in $\beta^2 \gamma$.  For our purpose, we will assume that
the fixed point reaction rate, $k^*/D = 2 \beta J - \beta^2 \gamma$, is
always positive.  Note that irrelevant details can renormalize (a finite
amount) all of the parameters of the model.  Dipole screening
leading to a dielectric constant greater than unity is an example of this
phenomenon.

We can again perform the matching.  Since at the fixed point the reaction
step is still  rate limiting \cite{GinzburgII},
 we find the same classical decay as
for the ion-pairing reaction:
\begin{equation}
\langle c_{\rm A}(t) \rangle
\sim \frac{1}{k^* t} \left( \frac{t}{t_0} \right)^\delta \ .
\label{13}
\end{equation}
Since reactant segregation is suppressed,
this result for the high-temperature ionic
reaction $\mbox{A}^+ +\mbox{B}^-
 \to \emptyset$
 is identical to that for
the neutral reaction 
$\mbox{A}+\mbox{A}
 \to \emptyset$
except for a different value of $k^*$
\cite{Deem1}.

\section{Conclusions}
There are many systems 
well-modeled by the 2-D Coulomb gas.
A simple physical system might be, for example,
ions confined
to a thin film between two insulators.
Other examples include dislocations or 
disclinations in systems such as charge density waves,
Abrikosov flux lattices, or Langmuir-Blodgett films.
In all cases, the defects unbind at
higher temperatures, in a form of
Kosterlitz-Thouless-Berezinskii  transition.
In the case of disclinations, or scalar charges, this transition is
exactly of the form that we consider, and the system is a perfect instance
of the 2-D Coulomb gas model.  The type of disorder that we
consider often comes about in these systems via pinning of some of the
defects.  The density of impurities, which are disrupting the
low-temperature phase, can be controlled via the number of
surface defects and is given roughly by
$\rho = \sqrt \gamma/J$.

For these systems we make the following experimental predictions.
There should be a continuously-variable
transition temperature in the presence of long-ranged, 
logarithmic-type disorder.    This type of disorder is naturally induced
by impurity phases in these systems.
This equilibrium behavior has, in fact,
 been seen in the melting of hexatic monolayers 
\cite{Zasadzinski} and hexatic charge density waves \cite{LieberI,LieberII},
where disclinations pinned by surface defects lead to a 
continuous lowering of
the hexatic-liquid transition temperature.
In other words, these experiments have shown that the order-disorder
transition can be driven either by increasing temperature or
by increasing disorder.  In terms of Figure 3, these
experiments crossed the transition line by increasing the disorder, 
{\em i.\ e.}\ by moving vertically upwards.
Ionic reactions, such as those considered
in \cite{Huber,Jang,Yurke,Ginzburg,Oshanin,Ispolatov,GinzburgII}, should
decay as $\langle c_{\rm A}(t) \rangle
\sim 1/(2 \beta J D t)$ at long times
in the absence of disorder.  In the presence of long-ranged,
logarithmic-type disorder
\cite{Fisher,Kravtsov1,Kravtsov2,Bouchaud1,Bouchaud2,Honkonen1,Honkonen2,Derkachov1,Derkachov2},
ions at finite density should pair in the low-temperature
phase according to Eq.\ (\ref{9}).  Finally, the concentration of ions
undergoing a bimolecular chemical reaction at high temperature
in this same type of disorder should
decay as Eq.\ (\ref{13}).

%
%
%
%
%

\section*{Acknowledgment}
It is a pleasure to acknowledge discussions with David Nelson.
This research was supported by the National Science Foundation
through grants CHE--9705165 and CTS--9702403.

\bibliography{react3}

\begin{thebibliography}{10}

\bibitem{Kosterlitz}
J.~M. Kosterlitz and D.~J. Thouless, J. Phys. C {\bf 6},  1181  (1973).

\bibitem{Berezinskii}
V.~L. Berezinskii, Z. Eksp. Teor. Fiz {\bf 61},  1144  (1971), [Sov. Phys.-JETP
  {\bf 34}, 610 (1971)].

\bibitem{nelsonII}
D.~R. Nelson,  in {\em Phase Transitions and Critical Phenomena}, edited by C.
  Domb and J. Lebowitz (Academic Press, New York, 1983), Vol.~7.

\bibitem{McCauley}
J.~L. McCauley, J. Phys. Chem. {\bf 10},  689  (1977).

\bibitem{Huberman}
B.~A. Huberman, R.~J. Myerson, and S. Doniach, Phys. Rev. Lett. {\bf 40},  780
  (1978).

\bibitem{Myerson}
R.~J. Myerson, Phys. Rev. B {\bf 18},  3204  (1978).

\bibitem{Ambegaokar}
V. Ambegaokar, B.~I. Halperin, D.~R. Nelson, and E.~D. Siggia, Phys. Rev. Lett.
  {\bf 40},  783  (1978).

\bibitem{AmbegaokarII}
V. Ambegaokar, B.~I. Halperin, D.~R. Nelson, and E.~D. Siggia, Phys. Rev. B
  {\bf 21},  1806  (1980).

\bibitem{Halperin}
B.~I. Halperin and D.~R. Nelson, J. Low Temp. Phys. {\bf 36},  599  (1979).

\bibitem{AmbegaokarIII}
V. Ambegaokar and S. Teitel, Phys. Rev. B {\bf 19},  1667  (1979).

\bibitem{Fisher}
D.~S. Fisher, M.~P.~A. Fisher, and D.~A. Huse, Phys. Rev. B {\bf 43},  130
  (1991).

\bibitem{Dorsey}
A. Dorsey, Phys. Rev. B {\bf 43},  7575  (1991).

\bibitem{Minnhagen}
P. Minnhagen, O. Westman, A. Jonsson, and P. Olsson, Phys. Rev. Lett. {\bf 74},
   3672  (1995).

\bibitem{Bormann}
D. Bormann, Phys. Rev. Lett. {\bf 78},  4324  (1997).

\bibitem{Ohta}
T. Ohta, Prog. Theo. Phys. {\bf 60},  968  (1978).

\bibitem{OhtaII}
T. Ohta and D. Jasnow, Phys. Rev. B {\bf 20},  139  (1979).

\bibitem{Knops}
H.~J.~F. Knops and L.~W.~J. den Ouden, Physica A {\bf 103},  597  (1980).

\bibitem{Amit}
D.~J. Amit, Y.~Y. Goldschmidt, and G. Grinstein, J. Phys. A {\bf 13},  585
  (1980).

\bibitem{Deem2}
M.~W. Deem and J.-M. Park, Phys. Rev. E {\bf 57},  2681  (1998).

\bibitem{Huber}
G. Huber and P. Alstrom, Physica A {\bf 195},  448  (1993).

\bibitem{Jang}
W.~G. Jang, V.~V. Ginzburg, C.~D. Muzny, and N.~A. Clark, Phys. Rev. E {\bf
  51},  411  (1995).

\bibitem{Yurke}
B. Yurke, A.~N. Pargellis, T. Kovacs, and D.~A. Huse, Phys. Rev. E {\bf 47},
  1525  (1993).

\bibitem{Ginzburg}
V.~V. Ginzburg, P.~D. Beale, and N.~A. Clark, Phys. Rev. E {\bf 52},  2583
  (1995).

\bibitem{Oshanin}
G.~S. Oshanin, A.~A. Ovchinnikov, and S.~F. Burlatsky, J. Phys. A {\bf 22},
  L977  (1989).

\bibitem{Ispolatov}
I. Ispolatov and P. Krapivsky, Phys. Rev. E {\bf 53},  3154  (1996).

\bibitem{GinzburgII}
V.~V. Ginzburg, L. Radzihovsky, and N.~A. Clark, Phys. Rev. E {\bf 55},  395
  (1997).

\bibitem{Kravtsov1}
V.~E. Kravtsov, I.~V. Lerner, and V.~I. Yudson, J. Phys. A {\bf 18},  L703
  (1985).

\bibitem{Kravtsov2}
V.~E. Kravtsov, I.~V. Lerner, and V.~I. Yudson, Phys. Lett. A {\bf 119},  203
  (1986).

\bibitem{Bouchaud1}
J.~P. Bouchaud, A. Comtet, A. Georges, and P.~L. Doussal, J. Phys {\bf 48},
  1445  (1987).

\bibitem{Bouchaud2}
J.~P. Bouchaud, A. Comtet, A. Georges, and P.~L. Doussal, J. Phys {\bf 49},
  369  (1988).

\bibitem{Honkonen1}
J. Honkonen, Y.~M. Pis'mak, and A.~V. Vasil'ev, J. Phys. A {\bf 21},  L835
  (1988).

\bibitem{Honkonen2}
J. Honkonen and Y.~M. Pis'mak, J. Phys. A {\bf 22},  L899  (1989).

\bibitem{Derkachov1}
S.~{\'E}. Derkachov, J. Honkonen, and Y.~M. Pis'mak, J. Phys. A {\bf 23},  L735
   (1990).

\bibitem{Derkachov2}
S.~{\'E}. Derkachov, J. Honkonen, and Y.~M. Pis'mak, J. Phys. A {\bf 23},  5563
   (1990).

\bibitem{Peliti}
L. Peliti, J. Phys. A {\bf 19},  L365  (1986).

\bibitem{Lee1}
B.~P. Lee, J. Phys. A {\bf 27},  2633  (1994).

\bibitem{Lee2}
B.~P. Lee and J. Cardy, J. Stat. Phys. {\bf 80},  971  (1995);
{\bf  87}, 951 (1997).

\bibitem{Deem1}
J.-M. Park and M.~W. Deem, Phys. Rev. E {\bf 57},  3618  (1998).

\bibitem{Cates}
M.~E. Cates and R.~C. Ball, J. Phys. (Paris) {\bf 49},  2009  (1988).

\bibitem{Wu}
D. Wu, K. Hui, and D. Chandler, J. Chem. Phys. {\bf 96},  835  (1992).

\bibitem{Nelson2}
D.~R. Nelson, Phys. Rev. B {\bf 27},  2902  (1983).

\bibitem{Fertig}
M.-C. Cha and H.~A. Fertig, Phys. Rev. Lett. {\bf 74},  4867  (1995).

\bibitem{Shraiman}
M. Rubinstein, B. Shraiman, and D.~R. Nelson, Phys. Rev. B {\bf 27},  1800
  (1983).

\bibitem{Zasadzinski}
R. Viswanathan, L.~L. Madsen, J.~A. Zasadzinski, and D.~K. Schwartz, Science
  {\bf 269},  51  (1995).

\bibitem{LieberI}
H.~J. Dai and C.~M. Lieber, Phys. Rev. Lett. {\bf 69},  1576  (1992).

\bibitem{LieberII}
H.~J. Dai, J. Liu, and C.~M. Lieber, Phys. Rev. Lett. {\bf 72},  748  (1994).

\end{thebibliography}

\end{document}